\documentstyle[revtex]{aps}
\begin{document}
\draft
\begin{title}
The neutron halo of $^6$He in a microscopic model
\end{title}
\author{Attila Cs\'ot\'o\cite{email}}
\begin{instit}
Institute of Nuclear Research of the Hungarian Academy of Sciences \\
P.O.Box 51 Debrecen, H--4001, Hungary
\end{instit}
\receipt{8 March 1993}
\begin{abstract}
The two--neutron separation energy of $^6$He has been reproduced
for the first time in a realistic parameter--free microscopic
multicluster model comprising the $\alpha +n+n$ and $t+t$
clusterizations, with $\alpha$ cluster breathing excitations
included. The contribution of the $t+t$ channel is substantial.
A very thick (0.85 fm) neutron halo has been
found in full agreement with the results of the latest
phenomenological analysis.
\end{abstract}


\section{Introduction}

Recently nuclei far from stability attract much interest in
nuclear physics.
Prominent representatives of these nuclei are e.g. $^6$He,
$^8$He, $^8$Li, $^8$B, $^{11}$Li. There are several calculations
for their description in macroscopic
\cite{Danilin,Kamimurali11,Kukulin,Lehman}, semimicroscopic
\cite{Suzuki} and microscopic \cite{Baye} models. In this paper
we use our dynamical microscopic
multiconfiguration multicluster model, developed and applied
recently to the ground
state of $^6$Li \cite{Csoto}, to study the neutron halo structure
of the ground state of $^6$He. All realistic
macroscopic three--body models underbind $^6$He by about 0.6--0.3
MeV. The situation is similar for $^{11}$Li, which is predicted
to be unbound by the most realistic parameter--free variational
calculation \cite{Kamimurali11}. Our first aim is
to check the validity of these models by comparing them with
our microscopic model, and to understand the physics of this
underbinding. Secondly, we calculate the thickness of the neutron
halo of $^6$He. For this quantity there are two contradicting
experimental predictions, both of them are based on Glauber--type
analyses of certain reaction cross section data. A simpler
analysis gives $\sim$ 0.4 fm \cite{Tanihata1}, while the other
one, which comes from more realistic model assumptions, results
in $\sim$ 0.9 fm \cite{Tanihata2}. The latter result is well
reproduced in a relativistic mean field model \cite{Tanihata2}.
As we shall see, our model strongly supports the second
prediction, too.

As the $\alpha$ particle is an inert cluster, it is natural to
assume that an $\alpha +n+n$ model is a most adequate one for the
description of $^6$He.
This nucleus is said to be borromean \cite{DanilinPR}, which means
that after the removal of any of the three clusters, the
remaining nucleus decays into two fragments. This indicates that
$^6$He has a genuine three--body nature and explains why the
macroscopic three--body models are
so successful in describing its ground state structure and
reactions \cite{DanilinPR}. Although several physical quantities
have been calculated, and a good overall agreement with
experiment has been reached in these models, it is neccessary to
investigate the validity of their foundation starting from
microscopic grounds, because nucleon
exchange and cluster rearrangement effects are expected to be
important in this mass range.

\section{Model}

The microscopic dynamical multiconfiguration
three--cluster model starts from the following trial function
for the six--body problem:
\begin{eqnarray}
\Psi
&=&\sum_{S,l_1,l_2,L}\Psi_{S,(l_1l_2)L}^{\alpha
(nn)}+\sum_{S,l_1,l_2,L}\Psi_{S,(l_1l_2)L}^{n (\alpha n)}=
\nonumber \\
&=&\sum_{S,l_1,l_2,L}\sum_i^{N_c}{\cal A}\left \{\left [ \left [\Phi
^\alpha_i(\Phi ^n\Phi^n)
\right ]_S
[\chi ^{nn}_{il_1}(\mbox{\boldmath $\rho $}_{nn})
\chi ^{\alpha (nn)}_{il_2}(\mbox{\boldmath $\rho $}_{\alpha (nn)})]
\raise-0.72ex\hbox{\scriptsize $L$}
\right ]_{JM}
\right \}
\nonumber \\
&+&\sum_{S,l_1,l_2,L}\sum_i^{N_c}{\cal A}\left \{\left [ \left
[\Phi ^n(\Phi ^\alpha_i\Phi^n)\right ]_S
[\chi ^{\alpha n}_{il_1}(\mbox{\boldmath $\rho $}_{\alpha n})
\chi ^{n (\alpha n)}_{il_2}(\mbox{\boldmath $\rho $}_{n(\alpha
n)})] \raise-0.72ex\hbox{\scriptsize $L$}
\right ]_{JM}
\right \}.
\label{wfn}
\end{eqnarray}
\noindent
Here ${\cal A}$ is the intercluster antisymmetrizer, the
\mbox{\boldmath $\rho $} vectors are the different intercluster
Jacobi coordinates, and [\ ] denotes angular momentum coupling.
While $\Phi ^n$ is a neutron spin--isospin eigenstate, $\Phi
^\alpha_i$ denote the ground state and some excited states of
the antisymmetrized $\alpha$ particle, and has the form
\begin{equation}
\Phi ^\alpha _i=
\sum_{j=1}^{N_c}A_{ij}\phi ^\alpha _{\beta _j},\ \
i=0,1,...,(N_c-1), \label{alpha}
\end{equation}
where $\phi ^\alpha _{\beta _j}$ is a translation invariant
shell--model state of the $\alpha$ particle with size
parameter $\beta _j$ and the $A_{ij}$ parameters are to be
determined by minimizing the energy of
the $\alpha$ particle \cite{Tang}.
In the literature the models using
such internal states for the description of the free clusters
are called distortion models \cite{Tang} or breathing cluster
models \cite{KBD}. This allows us to take into account the
distortion
of the $\alpha$ particle caused by the two outer neutrons. The
contribution of this distortion effect to the binding energy of
$^6$He yields information
about the role of the $\alpha$--breaking rearrangement channels.
Putting (\ref{wfn}) into the six--nucleon Sch\"odinger equation, we
arrive at an equation for the intercluster relative motion functions
$\chi$. These functions are expanded in terms of the so called
tempered Gaussian functions \cite{Kamimura} with different
ranges. E.g.\ in the case of $\chi ^{\alpha
n}_{i,lm}(\mbox{\boldmath $\rho $}_{\alpha n})$
\begin{equation}
\chi ^{\alpha n}_{i,lm}(\mbox{\boldmath $\rho $}_{\alpha
n})=\sum_{k=1}^{N_{\alpha n}}c_{ik}\Gamma_{lk}^{\alpha
n}(\mbox{\boldmath $\rho $}_{\alpha n}),
\label{chi}
\end{equation}
where
\begin{equation}
\Gamma_{lk}^{\alpha n}(\mbox{\boldmath $\rho $}_{\alpha n})=
\left ({{2^{l+1}(2\nu_k)^{l+3/2}}
\over{\sqrt{\pi}(2l+1)!!}}\right )
^{1/2}\rho_{\alpha n}^le^{-\nu_k\rho_{\alpha
n}^2}Y_{lm}(\widehat {
\mbox{\boldmath $\rho $}}_{\alpha n})
\label{tempgauss}
\end{equation}
and the coefficients $c_{ik}$ are to be determined from a
variational
principle with Eq.\ (\ref{wfn}) as a trial function. The use of
the tempered Gaussian functions makes it possible to
calculate all neccessary matrix elements analytically. The
present microscopic model is the same in many respects as was
used in \cite{Csoto} for the ground state of $^6$Li.

Thus we describe $^6$He as a superposition of the $\alpha (nn)$ and
$n(\alpha n)$ partitions with a distortable $\alpha$ particle being
in a spin--isospin zero state. The J$^\pi$=0$^+$ ground state of
$^6$He allows only the $(L,S)=(0,0)$ and $(1,1)$ states to
contribute. In the $[S,(l_1l_2)L]J$ coupling scheme we take into
account all important components that have any significance in
the binding energy. It is found that, if restricted to the
$\alpha +n+n$ space, from the point of view of the energy
it is enough to include the following terms in the trial
function:
\begin{equation}
\Psi=\Psi_{0,(00)0}^{\alpha (nn)}+
\Psi_{1,(11)1}^{\alpha (nn)}+
\Psi_{0,(00)0}^{n(\alpha n)}+
\Psi_{0,(11)0}^{n(\alpha n)}+
\Psi_{1,(11)1}^{n(\alpha n)}.
\label{Psi}
\end{equation}
Obviously, the role of the presence of the $n(\alpha n)$ type
clusterizations (second term on the rhs of (\ref{wfn})) is that in
the $\alpha (nn)$ configuration we can drop the higher angular
momentum components (e.g.\ 0,(22)0; 0,(44)0; ...) which
components, however, are non--negligible in macroscopic
three--body models, that only include the $\alpha (nn)$ type
configurations. This
finding has been confirmed by test calculations. In this paper
we use the same $\nu$ parameters in Eq.\ (\ref{tempgauss}) as were
used in \cite{Csoto} for the $N+N$ and $\alpha +N$ relative motions,
respectively. Several test calculations show, however, that, if
the model space is ``complete enough'', like in our case, the
results are independent of the choice of the basis parameters,
whenever the basis is well--balanced and spatially extensive enough.

\section{Interaction}

In microscopic calculations the choice of the effective $N-N$
interaction is a crucial point. If one wants a model to have
anything to do with the real physical problem in question, one
must be sure that the $N-N$ interaction is appropriate for all
subsystems which appear in the model. We must emphasize that
from this point of view the macroscopic approaches are in a more
favourable position. While in microscopic models the
interaction is strongly constrained by the state spaces assigned to
the cluster internal motions, which fact manifests itself in that
the bulk properties of the free clusters must be more or less
reproduced, in macroscopic approaches the binding energies and the
rms radii of the free clusters not even appear explicitly. Since
the $\{\alpha (nn);0,(00)0\}$ configuration is expected to have
a significant role in
$^6$He, one of the important subsystems we have to pay attention to
is the $^{2S+1}L_J=$$^1S_0$ singlet $N+N$ state. This is an
antibound state with negative energy but at the same time with
negative imaginary wave number \cite{Bethe}. It means that the
specification of this state by its energy is not sufficient. One
can distinguish between the bound and antibound states only by
an analysis of the scattering phase shift or by the
determination of the pole position of the $S$--matrix of the
$N+N$ scattering or by looking into the
effective--range parameters. Most of the published effective $N-N$
interactions in the cluster model literature do not take special
care of this state. Those that contain only space
exchanges (i.e.\ only Majorana parameter) in addition to a
Wigner force are definitely inappropriate because if they bind
the triplet deuteron, they bind the $^1S_0$ singlet dinucleon
states, too. That is they produce a bound singlet
dineutron which contradicts the borromean nature of $^6$He. A very
popular potential family of such kind are the Volkov forces
\cite{Volkov}.

There are few cases where definite attention has been paid to
this $^1S_0$ state during the construction of the interaction. In
\cite{Hackenbroich} the aim was to find an effective $N-N$ force
which gives good results for the two--nucleon systems comparable
with the results of realistic forces (e.g.\ \cite{Reid}). But at the
same time the quality of the description of other nuclei (e.g.\ the
$\alpha$ particle) with this force has not been investigated. We
have done a pilot calculation for the $\alpha$ particle
described by a single $0s$ shell model state which gives correct
rms radius for $\alpha$ and found it unbound by about 20 MeV.
The inclusion of a few $\alpha$ breathing modes makes the
situation better but the $\alpha$ particle still remains
unbound. The reason of this failure is obviously the fact that
this force would allow the $D$--state to play significant role
in the $\alpha$ cluster, but our model can handle only
$0s$--state clusters.

Another interaction is proposed by Tang {\em et al.\ }\cite{TLT}.
This reproduces the experimental $s$--wave $p+p$ effective--range
parameters, gives realistic results for the bulk properties of all
$0s$--shell nuclei and, supplemented by a spin--orbit force
\cite{Reichstein}, reproduces the $p$--wave $\alpha +N$
scattering phase shifts up to 15 MeV very
well. The form of this interaction is
\begin{eqnarray}
V_{ij}({\bf r})&=&\left [V_1+{{1}\over{2}}(1+P_{ij}^\sigma)V_2+
{{1}\over{2}}(1-P_{ij}^\sigma)V_3 \right ] \nonumber \\
&\times&\left
[{{1}\over{2}}u+{{1}\over{2}}(2-u)P_{ij}^r \right ] \nonumber \\
&+&{{1-\tau_{iz}}\over{2}}{{1-\tau_{jz}}\over{2}}{{e^2}
\over{r_{ij}}}+V_4\hbar^{-2}{\bf l}(\mbox{\boldmath $\sigma
$}_i+\mbox{\boldmath $\sigma $}_j),
\label{pot}
\end{eqnarray}
where $P^r$ and $P^\sigma$ are the space-- and spin--exchange
operators, respectively, u is the exchange mixture parameter,
${\bf r}={\bf r}_2-{\bf r}_1$,
$\tau_{iz}$ are the $z$ components of the isospin vector, {\boldmath
$\sigma $}$_i$ are the Pauli vectors of the nucleonic spin, and
\begin{equation}
{\bf l}=-{\textstyle {1\over 2}} i\hbar {\bf r}
\times (\nabla _2-\nabla _1),
\end{equation}
is the orbital momentum of the relative motion of the two
nucleons. The potentials have the following Gaussian form factors:
\begin{equation}
V_k=V_{0k}\exp (-r_{ij}^2/a_k^2),\ \  k=1,...,4.
\end{equation}

There are two shortcomings of the Tang force that we should
mention. One of them comes from the fact that the description of
the two--nucleon states is good only in an effective manner,
i.e.\ the phase shifts are not perfect and in the triplet $p+n$
case the experimental deuteron energy is reproduced by assuming
a pure $^3S_1$ state. The allowance of the coupling between the
$^3S_1$ and $^3D_1$ states by the inclusion of a tensor force
ruins the picture because it considerably lowers the binding
energy of the deuteron. The second problem is that, as a result
of the lack of
tensor force, the anomalous low energy 1,2,0 level order of the
$^3P_J$ ($J=0,1,2$) states \cite{Arndt} cannot be reproduced.
As far as a $J^\pi =0^+$ state is concerned in the six--nucleon
system, the first shortcoming does not cause any problem because
in this case a triplet $S$ state between the two outer nucleons
cannot be present. However it can give rise to difficulties,
e.g.\ in the description of the $1^+$ ground state of
$^6$Li. Our test
calculations show that this really happens. In a model,
which contains all possible angular momentum configurations of any
significance this force overbinds the ground state of $^6$Li by
about 1.2 MeV. To check the effect of the second shortcoming, we
supplemented the Tang force by a tensor term due to Heiss and
Hackenbroich \cite{Heiss}; this combination gave good results
for the five--nucleon systems \cite{Langanke}. We have
determined its exchange mixture parameters so as to reproduce the
anomalous $^3P_J$ $N+N$ level splitting qualitatively. Since our
tensor force is based only on this not very firm ground, we do not
use it in $^6$He except for test purposes.

Our force parameters are as follows:
\begin{eqnarray}
V_{01}=\ \ 200.0 \mbox{ MeV}, \ \ \ a_1=0.82006 \mbox{
fm},\nonumber\\
V_{02}=-178.0 \mbox{ MeV}, \ \ \  a_2=1.25098 \mbox{
fm},\nonumber\\
V_{03}=-91.85 \mbox{ MeV}, \ \ \  a_3=1.46647 \mbox{
fm},\nonumber\\
V_{04}=-591.1 \mbox{ MeV}, \ \ \  a_4=0.57735 \mbox{
fm}.\nonumber
\end{eqnarray}
We have chosen $N_C$ in (\ref{alpha}) to be 3, which yields the
following size parameters: $\beta_1$=0.355 fm$^{-2}$,
$\beta_2$=0.795 fm$^{-2}$, $\beta_3$=2.66 fm$^{-2}$. The energy
and the point matter rms radius of the $\alpha$ particle is
--25.60 MeV and 1.407 fm, respectively to be compared to the
experimental values --28.296 MeV and 1.48 fm. The exchange
mixture parameter $u$ is set to be 0.98 to fit the
$\alpha +N$ phase shifts as are seen in Fig.\ \ref{fig1}
(further on we
show the $\alpha$+n and $p+p$ scattering phase shifts). Although the
theoretical effective--range parameters of the $^1S_0$ $N+N$
state are close to the experiment, the phase shift is not
perfect, but the antibound character is very well reproduced
(Fig.\ \ref{fig2}). It is easy to see that
in the $\alpha (nn)$ configuration only the $^3P_1$ triplet--odd
$N+N$ state
can be present in the $J^\pi=0^+$ state of the six--nucleon
system. In this partial wave the agreement with experiment is
also fairly good, although the inclusion of the tensor force
renders
the theoretical results closer to the experiment.
It will be shown
that higher partial waves both in the $N+N$ and $\alpha +N$
relative motions can safely be neglected. We emphasize that all
parameters are fixed to independent data, that is, as far as
$^6$He is concerned, our model is parameter--free.

\section{Results and discussion}

The above interaction sets the two--neutron separation energy of
the $^6$He g.s.\ at 0.740 MeV, which is to be compared to the
experimental value 0.975 MeV \cite{Ajzenberg}. The weights of the
$L=0$, $S=0$ and $L=1$, $S=1$ components are in Table \ref{tab1},
along with the results of \cite{Danilin},
\cite{Kukulin} and \cite{Lehman}. To quantify the roles of the
different non--orthogonal channels we
have computed the amounts of clustering $S_\mu$ of the different
clusterizations $\mu$. This quantity is the weight of the
component of the wave function $\Psi$ that lies in the segment
of the state space associated with clusterization $\mu$
\cite{AP}:
\begin{equation}
S_\mu=\left\langle\right . \Psi \left \vert \right . P_\mu
\left \vert \right . \Psi \left .\right\rangle
\end{equation}
where $P_\mu$ projects onto the subspace $\mu$. Here we used the
method of \cite{Csoto} to calculate $S_\mu$. The resulting
values are in Table \ref{tab2}, while the contribution of the
different channels to the energy of $^6$He is in Table \ref{tab3}.
It is interesting to see that the amount
of clustering of the $\{n(\alpha _0n); 0,(00)0\}$ component is
larger than that of $\{\alpha _0(nn); 0,(00)0\}$ in spite of the
fact that the latter contributes much more to the binding energy
than the former. This supports
that in the pure $\alpha (nn)$ type macroscopic models there are
non--negligible components with higher angular momenta (e.g.
$\{\alpha (nn); 0,(22)0\}$). In our model these components are
well represented by the $n(\alpha n)$ type components. To check
this we complemented our model space (\ref{Psi}) by
$\Psi_{0,(22)0}^{\alpha (nn)}$ and $\Psi_{0,(22)0}^{n(\alpha
n)}$. The energy gain is really small, being 0.0025 MeV and the
amounts of these clusterizations are 0.035 for $\{\alpha_0 (nn);
0,(22)0\}$ and 0.015 for $\{n(\alpha_0 n); 0,(22)0\}$.

For further test purposes we have done calculations for the two
other members of the $J^\pi=0^+$, $T=0$ isospin triplet in
$A=6$, namely on $^6$Li and $^6$Be. Since the g.s.\ of $^6$Be is
unbound, our present method for $^6$Be is a pseudobound--state
approximation \cite{LovasNag}. In $^6$Be the same
components are taken into account as in Eq.\ (\ref{Psi}), while in
$^6$Li all $\alpha (pn)$, $p(\alpha n)$ and $n(\alpha p)$
clusterizations are present and in $\alpha (pn)$ we include the
$0,(11)0$ component as well. The presence of this $0,(11)0$
component requires the $^1P_1$ $N+N$ subsystem to be correct. We
checked that our
phase shift is close to experiment in this partial wave, too.
Table \ref{tab1} shows that the deviations from experiment are
much larger in \cite{Kukulin} than in our case, but our results
agree well with the results of \cite{Danilin}. (In
\cite{Danilin} no parameter--free results were published for
$^6$Li and $^6$Be, but the estimation of the Coulomb energies
gave results similar to ours for these nuclei, too.) The
situation is similar in respect of the weights, our results being
closer to the results of \cite{Danilin}. In contrast with
\cite{Danilin} however, in our case the weights of the
$(L,S)=(1,1)$ component are decreasing from $^6$He to $^6$Be.
The origin of this contradicting behaviour in microscopic and
macroscopic models is not known. The agreement between our
results and the results of the best macroscopic model
\cite{Danilin} both with respect to the separation
energies and the weights of the various $(L,S)$ components is
remarkably good. Since this macroscopic model is also
parameter--free and the description of the subsystems has similar
quality (in \cite{Danilin} the
description of the $N+N$ states is almost perfect but the
$\alpha +N$ phase shifts are worse than ours), this agreement
strongly supports the validity of \cite{Danilin}--type macroscopic
three--body approaches in describing $^6$He.

Having found our model appropriate for $^6$He from the point of view
of both energetic and fragmentation properties, the next step is to
calculate the thickness of its neutron halo.
We have calculated the point nucleon root mean square matter,
neutron and proton radii of $^6$He. The results are presented in
Table \ref{tab4}. Although our $\alpha$ particle is a bit
smaller than it should be, which implies that
the resulting radii of $^6$He may be smaller by the same amount,
the thickness of the neutron halo is not affected by this dicrepancy.
In our present theoretical model the neutron halo is
considerably thicker than the one which was extracted from
certain reaction cross section measurements \cite{Tanihata1},
but is in a very good agreement with the result of the latest
phenomenological analysis \cite{Tanihata2}. The agreement with
\cite{Suzuki}, where a cluster orbital shell model was used for
$^6$He is also remarkable. It is known, however, that the radii
trivially depend on the separation energy \cite{FragProp}, and
our separation energy is not correct. That is why we have to seek
further after the possible origin of this energy lack.

The same amount of binding
energy lack in microscopic and macroscopic models excludes the
possibility that the same mechanism causes the lack of
$^6$He binding energy as causes the lack of binding energy of
the $\alpha$ particle itself, in microscopic models. To see the
role of the breathing modes of the $\alpha$ particle, we repeated
the calculations with $N_c$=1 in (\ref{wfn}). In this case the
energy optimized $\beta$ is 0.606 fm$^{-2}$. The $\alpha +N$
phase shifts remain
almost unchanged (in Fig.\ \ref{fig1} they are indistinguishable
from the
phase shifts of the $N_c=3$ case) but the two--neutron separation
energy becomes 0.652 MeV, i.e.\ the distortion of the $\alpha$
particle results in a roughly 0.09 MeV gain in $^6$He. The fact
that the $\alpha$ breathing modes hardly affect the $\alpha +N$
scattering but substantially contribute to the energy of $^6$He
makes it probable that the $\alpha$ breathing modes represent
not a two--body on--shell excitation effect, but a three--body
off--shell excitation--deexcitation mechanism, i.e.\ the
core polarization of the $\alpha$ particle. A similar effect has
been found in the g.s.\ of $^6$Li \cite{Csoto}. Of course, our
finding is not a proof of the presence of core polarization, it
should be investigated by further calculations. We mention here
that the probable role of core polarization in $^{11}$Li was
emphasized in \cite{Kamimurali11}.
The importance of $\alpha$ breathing modes
indicates that the $\alpha$--beakup rearrangement channels may
have large weights in $^6$He. This is in agreement with the
finding of \cite{Wiesb}, where the important role of the
$t+$$^3$He rearrangement channel has been pointed out in the g.s.\ of
$^6$Li. The most likely of these
channels in $^6$He is the $t+t$ clusterization. As we can see in
Table \ref{tab2}, the weight of this clusterization is large (0.496)
even in a model space which does not contain explicitly a $t+t$
term.

Therefore we have supplemented our model space (\ref{Psi}) by a
$t+t$ term
\begin{equation}
\Psi _{0,0}^{tt}
={\cal A}\left \{\left [ [\Phi ^t\Phi ^t
]\raise-0.66ex\hbox{\scriptsize 0}
\chi ^{tt}_0(\mbox{\boldmath $\rho $}_{tt})
\right ]_{00} \right \}.
\label{tt}
\end{equation}
The energy optimized size parameter of the triton is
$\beta_t=0.451$ fm$^{-2}$, which gives --4.56 MeV and 1.48 fm
for the energy and radius of $t$, respectively to be compared to
the experimental values --8.482 MeV and 1.49 fm.
The inclusion of (\ref{tt}) into the trial function puts the
two--neutron separation energy of $^6$He from 0.740
MeV to 1.416 MeV, that is $^6$He becomes overbound by about 0.44
MeV. This can be explained by recalling that the effective force
tailored for a state space that does not include rearrangement
channels is more restrictive than the $\{\alpha
+n+n;t+t\}$ state space. To cure this shortcoming we have
repeated the calculations of the $\alpha +n$ phase shifts in
a $\{\alpha +n;d+t\}$ model space (all possible
$(L_{dt},S_{dt})$ components are included except the (3,$3\over
2$) one in the ${{3}\over{2}}^-$ state of $^5$He, because of the
lack of the tensor force). It turned out that the choice of the
deuteron internal state is a crucial point. To be consistent
with the triton we have chosen only one size parameter for the
deuteron, too. The deuteron appearing in the five--nucleon system
is supposed to compose a triton inside $^6$He with the additional
neutron. This involves a strong restriction for the
size parameter of the deuteron. To see this, we express the
internal state of triton,
\begin{equation}
\Phi ^t=\exp\left (-{{\beta_t}\over{2}}\sum_{i=1}^3({\bf
r}_i-{\bf R}_t)^2\right )
\end{equation}
(where ${\bf R}_t$ is the triton center
of mass) in terms of the ${\bf r}_d={\bf r}_{n_1}-{\bf r}_p$,
${\bf r}_{nd}={\bf r}_{n_2}-({\bf r}_{n_1}+{\bf r}_p)/2$ and ${\bf
R}_t$ coordinates (where $n_1$ and $n_2$ denote the two neutrons)
\begin{equation}
\Phi ^t=\exp\left (-{{\beta_t}\over{2}}(3{\bf R}^2_t+{{2}\over{3}}
{\bf r}^2_{nd}+{{1}\over{2}}{\bf r}^2_d)\right ).
\end{equation}
After this, it is obvious that the deuteron size parameter must
be chosen as $\beta_t/2$.

We have fitted two parameters of our interaction ($u$=0.92,
$V_4$=--691.1 MeV) so as to get results for the $\alpha +N$
scattering with almost the same quality as in the one--channel
case (Fig.\ \ref{fig1}). (In the ${{1}\over{2}}^+$ and
${{3}\over{2}}^-$
partial waves the difference from the pure $\alpha +n$ results
are within the line thickness.)
The change of the mixing parameter $u$ slightly changes the $^3P_1$
and $^1P_1$ $N-N$ phase
shifts as well, but these changes are all well within 0.4 degree.
Using this readjusted force in the $\{\alpha +n+n;t+t\}$ model
space, we get 0.961 MeV for $^6$He which is very
close to the experimental value, 0.975 MeV. To check the inherent
uncertainity connected with the poorer fit of the ${{1}\over{2}}^-$
$\alpha +n$ phase shift we have selected another $(u,V_4)$
combination (0.94,--641.1 MeV) which produces the dotted line in
Fig.\ \ref{fig1}. This force gives 1.012 MeV for $^6$He; that is the
uncertainity is as small as 0.05 MeV. The amount of $t+t$
clustering is found to be 0.547, which confirms its major role in
$^6$He. The model spaces of $^6$Li and $^6$Be have also been
extended by the inclusion of a $t+$$^3$He and a $^3$He+$^3$He
component, respectively.
As we can see in Table \ref{tab1}, the two--proton separation
energy of $^6$Be also comes very close to the experiment,
however in $^6$Li approximately a 0.1 MeV is still missing. One
may think that our wave function is spatially not extensive
enough for such a loosely bound
state. To check this, we substantially extended the spatial
region of our trial function, almost doubling the number of
basis functions, but this yields only 0.008 MeV. This confirms
that our basis is well--balanced and the calculations are
numerically stable. The cause of the 0.1 MeV energy discrepancy is
probably that our force reproduces just the $p+p$ singlet s--wave
effective--range parameters. It was shown
\cite{Hackenbroich,Reid} that such a
force could reproduce the $n+n$ effective--range parameters, too,
but at the same time fails to reproduce the $p+n$ ones. In
\cite{Lehman} it was found that the change of an $N-N$ force,
which reproduces the $n+n$ data to another one which reproduces
the $p+n$ ones results in a $\sim$ 0.18 MeV decrease in the
two--neutron separation energy of $^6$He. This is in agreement
with our finding.

We have
recalculated the various radii using the readjusted force and
got the results as shown in Table \ref{tab4}. We can see that
the 0.2 MeV increase in the $^6$He two--neutron separation
energy only slightly modifies the radii,
that is the neutron halo is rigid against the significant
modification of the model. Finally we have carried out a
calculation including the tensor interaction. The result shows
that a tensor force which tends to
restore the anomalous level order in the $^3P_J$ states of the
two--nucleon system, lowers the two--neutron separation energy
of $^6$He. In our case this decrease is less than 0.08 MeV. We
should note, however,
that our $2E_t- E_\alpha$ threshold splitting is roughly 5 MeV
larger than the experimental value. This means that if this energy
were right, the weight of the $t+t$ clusterization would be
higher, which would increase the two--neutron separation energy
of $^6$He again.
This increasing would hopefully compensate the effect of the
tensor force.

\section{Conclusion}

In summary, we have done careful microscopic multiconfiguration
multicluster calculations for $^6$He. We have fixed all
parameters of our model to independent data, and the description
of all subsystems have been found reasonably good. In respect of
$^6$He our model is free of any parameter. Although our results
confirm the validity of the macroscopic three--body models, we have
found that the inclusion of some excited states of the $\alpha$
particle is significant and the $t+t$ rearrangement channel has
large weight even in a model which is almost complete in the $\alpha
+n+n$ model space. The $\alpha$ breathing modes and the $t+t$
clusterization are found to be responsible for the
missing binding energy of the macroscopic models. These effects are
expected to play important role in other halo nuclei, e.g.\ in
$^{11}$Li. Finally we have
found a very thick neutron halo in $^6$He which is in full
agreement with the latest experiments \cite{Tanihata2}. We have found
that this thickness is hardly sensitive to the model assumptions.
Of course, the novel features of our model can really
manifest their relevance in the calculation of various physical
quantities (electric dipole strength, $\beta$-decay branching
ratio, momentum distribution). These calculations as well as the
application of our model to some of the above--mentioned nuclei
are in progress.

\acknowledgments

This research was supported by OTKA (National Science Research
Foundation, Hungary) under contract numbers 3010 and F4348. The
author is indebted to Prof. R.~G. Lovas, Prof. Y. Suzuki and
Dr. K. Varga for stimulating discussions and to Prof. B.
Gyarmati for reading the manuscript.

\figure{Phase shifts for $\alpha +n$
scattering in $l=0$ and 1 states (dashed line: in the $\{\alpha
+n+n;t+t\}$ model with $u=0.92$, $V_4=-691.1$ MeV; dotted line:
the same model but with $u=0.94$, $V_4=-641.1$ MeV).
Experimental data are taken from \cite{Bond}.\label{fig1}}
\figure{$^1S_0$ phase shift of the $p+p$
scattering. Experimental data are taken from
\cite{Arndt}.\label{fig2}}

\newpage
\widetext
\begin{table}
\caption{Energies (relative to the $\alpha$ energy) and weights
of the $(L,S)$
components of the $J^\pi=0^+$, $T=1$ isospin triplet in $A=6$.}
\begin{tabular}{lcccr@{}lccccc}
&\multicolumn{3}{c}{$^6$He
(0$^+$,T=1)}&\multicolumn{4}{c}{$^6$Li (0$^+$,T=1)}&
\multicolumn{3}{c}{$^6$Be (0$^+$,T=1)} \\ \cline{2-4}
\cline{5-8} \cline{9-11}
Model&E (MeV)& (0,0) & (1,1) & \multicolumn{2}{c}{E (MeV)}&
(0,0) & (1,1) &
E (MeV)& (0,0) & (1,1) \\
\hline
Kukulin--86 \cite{Kukulin}& -- 0.138& 95.70 & \hphantom{1}4.30 &
0.&742& 96.30& \hphantom{1}3.70 & 2.083 &  ---  &  --- \\
Kukulin--92 \cite{Kukulin}& -- 0.025& 91.74 & \hphantom{1}8.26 &
0.&844& 92.75& \hphantom{1}7.25 & 2.102 & 94.89 &
\hphantom{1}5.11 \\
Danilin \cite{Danilin}& -- 0.731& 85.76& 14.71 & &---$^{\rm a}$&
84.75 &  15.26 & ---$^{\rm a}$ & 82.33 & 18.40 \\
\{$\alpha+N_1+N_2$\}$^{\rm b}$&-- 0.740& 86.24 & 13.76 & 0.&165
& 87.07 & 12.93 & 1.516 &
87.55  & 12.46 \\
\{$\alpha+N_1+N_2;T_1+T_2$\}$^{\rm c}$& -- 0.961& 88.15 & 11.85
& -- 0.&025& 88.75 & 11.25
& 1.357 & 89.08 & 10.92 \\
Experiment \cite{Ajzenberg}& -- 0.975&   ---   &   ---  & --
0.&137   &   ---    &  --- & 1.371  &   --- &   --- \\
\end{tabular}
\tablenotes{$^{\rm a}$No parameter--free data are available.}
\tablenotes{$^{\rm b}$For $^6$He: $N_1=N_2=n$; for $^6$Li:
$N_1=n,\ N_2=p$; for $^6$Be: $N_1=N_2=p$.}
\tablenotes{$^{\rm c}$For $^6$He: $T_1=T_2=t$; for $^6$Li:
$T_1=t,\ T_2=$$^3$He; for $^6$Be: $T_1=T_2=$$^3$He.}
\label{tab1}
\end{table}
\mediumtext
\begin{table}
\caption{Cluster decomposition of $^6$He [the three numbers in
each groups
are for the three $\alpha$ states].}
\begin{tabular}{ccl@{}l@{}ll@{}l@{}l}
\multicolumn{2}{c}{Clusterization}
                        & \multicolumn{6}{c}{Amount of clustering}\\
Partition & $S,(l_1l_2)L$ & \multicolumn{3}{c}{\{$\alpha+n+n$\}}
&\multicolumn{3}{c}{ \{$\alpha+n+n;t+t$\}}\\
\tableline
$\alpha (nn)$ & 0,(00)0
&(0.8378,&0.4457,&0.0004)&(0.8446,&0.4651,&0.0006) \\
$\alpha (nn)$ & 1,(11)1
&(0.1265,&0.0004,&0.00003)&(0.1086,&0.0003,&0.00004) \\
$n(\alpha n)$ & 0,(00)0
&(0.8443,&0.4696,&0.0006)&(0.8491,&0.4893,&0.0008) \\
$n(\alpha n)$ & 0,(11)0
&(0.8424,&0.0098,&0.0002)&(0.8469,&0.0153,&0.0003) \\
$n(\alpha n)$ & 1,(11)1
&(0.1363,&0.0005,&0.00006)&(0.1173,&0.0005,&0.00008) \\
$tt$ & 0,0 &
\multicolumn{3}{c}{0.4964}&\multicolumn{3}{c}{0.5471} \\
\end{tabular}
\label{tab2}
\end{table}
\narrowtext
\begin{table}
\caption{Change of $^6$He energy (relative to the
$\alpha$ energy) in MeV when each one of the clusterization
components is omitted.}
\begin{tabular}{cccc}
\multicolumn{2}{c}{Omitted component}  &
\multicolumn{2}{c}{Model} \\
Partition     & $S,(l_1l_2)L$ & \{$\alpha+n+n$\} &
\{$\alpha+n+n;t+t$\} \\
\tableline
\multicolumn{2}{c}{None}      & -- 0.740 & -- 0.961  \\
$\alpha (nn)$ &  0,(00)0      & -- 0.429 & -- 0.638  \\
$\alpha (nn)$ &  1,(11)1      & -- 0.708 & -- 0.924  \\
$n(\alpha n)$ &  0,(00)0      & -- 0.736 & -- 0.957  \\
$n(\alpha n)$ &  0,(11)0      & -- 0.643 & -- 0.876  \\
$n(\alpha n)$ &  1,(11)1      & -- 0.294 & -- 0.488  \\
$tt$          &  0,0          & ---      & -- 0.315  \\
\end{tabular}
\label{tab3}
\end{table}
\mediumtext
\begin{table}
\caption{Point nucleon rms radii (matter(m), neutron(n) and
proton(p)) and the thickness of the neutron halo in $^6$He (in fm).}
\begin{tabular}{lr@{}lr@{}lr@{}lr@{}l}
Model&&$r_m$&&$r_n$&&$r_p$&\multicolumn{2}{c}{$r_n-r_p$} \\
\tableline
Suzuki \cite{Suzuki}& 2&.40&2&.64&1&.82&0&.82 \\
\{$\alpha+n+n$\}&2&.440&2&.707&1&.7790&0&.917 \\
\{$\alpha+n+n;t+t$\}&2&.397&2&.648&1&.793&0&.855 \\
Experiment \cite{Tanihata1}&2&.48$\pm$ 0.03&2&.61 $\pm$ 0.03 &
2&.21 $\pm$ 0.03 & 0&.4 \\
Experiment
\cite{Tanihata2}&2&.33$\pm$0.04&2&.59$\pm$0.04&1&.72
$\pm$0.04&0&.87$\pm$0.06 \\
\end{tabular}
\label{tab4}
\end{table}
\end{document}